# Light emission in nanogaps: overcoming quenching

Jianji Yang, Rémi Faggiani, and Philippe Lalanne*

Laboratoire Photonique Numérique et Nanosciences, Institut d'Optique d'Aquitaine, Université Bordeaux, CNRS, 33405 Talence, France

* E-mail: philippe.lalanne@institutoptique.fr

**Abstract:** Very large spontaneous-emission-rate enhancements (~1000) are obtained for quantum emitters coupled with tiny plasmonic resonance, especially when emitters are placed in the mouth of nanogaps formed by metal nanoparticles that are nearly in contact. This fundamental effect of light emission at subwavelength scales is well documented and understood as resulting from the smallness of nanogap modes. In contrasts, it is much less obvious to figure out whether the radiation efficiency is high in these gaps, or if the emission is quenched by metal absorption especially for tiny gaps a few nanometers wide; the whole literature only contains scattered electromagnetic calculations on the subject, which suggest that absorption and quenching can be kept at a small level despite the emitter proximity to metal. Thus through analytical derivations in the limit of small gap thickness, it is our objective to clarify why quantum emitters in nanogap antennas offer good efficiencies, what are the circumstances in which high efficiency is obtained, and whether there exists an upper bound for the maximum efficiency achievable.

Spontaneous emission remains at the core of the performance of many optoelectronic devices, including not only lighting components and displays, but also lasers, optical amplifiers, single photon sources and non-classical light sources in general. Metal nanogaps formed by a thin insulator layer sandwiched between two metals films have very rich physical properties and many established applications ranging from electron tunneling microscopy, nanocatalysis, Raman spectroscopy to disruptive electronics, but they are also likely to profoundly impact spontaneous emission [1]. Owing to the strong localization in the gap, metal nanogaps strongly modify the electromagnetic density of modes. It follows that the spontaneous emission of dye molecules or quantum dots which are placed in the gap can be enhanced considerably. This fundamental phenomenon of light emission, known as the Purcell effect [2], has been first demonstrated in optics by coupling quantum emitters with resonant dielectric microcavities [3] with very high quality factors and mode volumes of the order of the wavelength cube. The use of deep-subwavelength confinements with plasmonic nanostructures has created a totally new framework with mode volumes 10,000 times smaller and broadband responses [4,5], and thus have opened promising route toward new applications in optical spectroscopy [6-8], spaser or low-threshold nanolasers [9-10], or broadband non-classical light sources [5].

For instance, for a molecule placed close to a single metallic nanoparticle, such as a nanorod or preferably a triangular particle with sharp corners, spontaneous emission rate enhancements of a few hundred are observed over a spectral linewidth of about a tenth of the emitted frequency. This is remarkable and unfeasible with dielectric structures. The down-side is that metal absorbs. High radiation efficiencies are achieved as long as the coupling with the particle resonance dominates, but as the molecule approaches the metal surfaces down to separation distances smaller than 10 nm, the emission efficiency breaks down. Photon emission is quenched. Quenching has been considered for many years as the predominant spontaneous decay channel for an emitter that is placed at a small separation distance *d* from a metallic object [11]. In classical electrodynamics, this effect, which usually scales as $\propto (k_0 d)^{-3}$ [12], is due to the intense near field of the emitter that induces considerable Ohmic heating of the metal. Figure 1 illustrates the textbook example of a vertically-polarized molecule (assumed to have an inherent quantum yield of 100% and treated as an electric dipole) emitting visible light at a small distance from a silver film. As the distance decreases below 5-10 nm, quenching becomes the dominant decay channel, and its rate far exceeds the other decay rates into photons and surface plasmon polaritons. A key reason is that, in contrast with the couplings of the emitter with surface plasmons and free-space modes, quenching is a highly-localized near-field effect that varies at the nanometer scale.

Since the very early stage of plasmonic-nanoantenna research, there was a concern that large spontaneous-emission-rate enhancements with metallic nanostructures would be inevitably accompanied by a strong quenching that would critically restrict antenna efficiencies. However, recent experiments [1,13-15] performed with molecules placed in nanogaps formed by pairs of particles placed side by side, such as patches [1], bowties [13], and nanoparticle dimmers [14,15] have shown that nanogaps offer enhancements even stronger than those achieved with single nanoparticles due to the capacitive coupling, and most importantly have revealed that the initial intuition is wrong. Recent experimental results obtained for nanocube-antenna [1] provide a particularly striking example. In the experiment, a significantly large Purcell factor of $10^3$ is measured and a good extraction efficiency of 50% is surprisingly predicted for dye molecules in an 8-nn-thin polymer-film sandwiched between a gold substrate and a small silver nanocube. Notably, the good efficiency is obtained for molecules that are placed only 4-nm away from the metal interfaces. At such small distances, quenching is considerable, see Fig. 1. Clearly another decay channel with a scaling dependence at least as large as $(k_0 d)^{-3}$ is available in metal-insulator-metal (MIM) nanogaps, and this deserves special studies.

To understand the physical process at play in light emission in tiny nanogaps, it is easier to consider Fig. 2 that sketches a zoomed view of the mouth of an MIM nanogap and outline the different processes at play in light emission. Primarily, the emitter may quench by creating electron-hole pairs directly into the metal with a rate denoted by $\gamma_{quench}$. Direct decay into free-space photons is likely to be negligible for tiny gaps, so that the second alternative decay channel is provided by the excitation of gap plasmons with a rate $\gamma_{GSP}$, which are available at the antenna mouth. Subsequently, the gap plasmons may heat the metal with a rate $\gamma_{abs}$, and since they are coherent electronic oscillations, they may also serve as a relay to further scatter into free-space photons with a rate $\gamma_{rad}$. Thus for an emitter with an inherent quantum yield of 100%, the antenna efficiency $\eta$ can be written as the product of two terms

$$\eta = \eta_e \, \eta_i \, . \quad (1)$$

The first term, $\eta_e = \gamma_{rad} / (\gamma_{abs} + \gamma_{rad})$ is dominantly impacted by the antenna capability to convert the energy mediated by gap plasmons into photons. Nanoantennas with progressive widening of the gap, such as bowties or sphere dimers, adiabatically taper gap plasmons into photons and are likely to offer nearly-perfect conversions. Conversely antennas with abrupt gap terminations, such as nanocube patch antennas that considerably boost the spontaneous decay rate because of the plasmonic cavity resonance [1], favor the absorption channel to the detriment of the radiation one. Thus the first term $\eta_e$ primarily refers to an extrinsic property that is optimized by engineering the antenna geometry. In contrast, the second term $\eta_i = \gamma_{GSP} / (\gamma_{GSP} + \gamma_{quench})$ is intrinsic as it only depends on the nature of nanogap, irrespectively of the precise shape of the antenna.

From Eq. (1), it becomes clear that high efficiency requires not just appropriate engineering of the MIM mouth ($\eta_e \approx 1$), but especially that the decay rate into gap plasmons overcomes the quenching decay rate ($\eta_i \approx 1$). Actually by using a complex continuation technique to calculate the Green-tensor, it can be shown [16] that the decay rate into gap plasmon scales as $(k_0 d)^{-3}$, exactly like quenching. Since $\gamma_{GSP}$ and $\gamma_{quench}$ have identical scaling, the efficiency $\eta_i$ becomes independent of *d* for small *d*'s and takes a simple expression that only depends on the material dielectric constants

$$\eta_i = \left( 1 + \frac{\varepsilon''}{2\varepsilon_d} \frac{|\varepsilon|^2}{|\varepsilon + \varepsilon_d|^2} \right)^{-1}, \quad (2)$$

where $\varepsilon(\omega) = \varepsilon' + i\varepsilon''$ and $\varepsilon_d$ are the relative permittivities of the metallic mouths and dielectric gap. The fact that $\gamma_{GSP}$ and $\gamma_{quench}$ have an identical rate-dependence with *d* is not trivial but should not come as a surprise; for vanishing $d$'s, gap plasmons exhibit slower group velocities ($v_g \prec d^{-1}$) and become mostly electronic waves with a low photonic character; they lose their delocalized coherent character, so that direct near-field absorption and gap-plasmon excitations contribute similarly to the local density of states seen by the quantum emitter. This intuition is quantitatively confirmed by the computational results of Figs. 3(a-b), which shows similar trends for $d < 10$ nm for quenching and gap-plasmon excitation rates of electric dipoles emitting in the center of a planar nanogap for the same dipole-orientation, materials and emission wavelength as those used for the single interface in Fig. 1.

It is interesting to consider Eq. (2) for $\varepsilon_d << -\varepsilon'$, e.g. for good metals, $\eta_i \approx \left(1+\varepsilon''/2\varepsilon_d\right)^{-1}$. The expression evidences that the intrinsic efficiency is enhanced for MIM nanogaps with high-index insulators and low loss metals. Direct application of the previous formula for semiconductor gaps ($\varepsilon_d = 12$) gives an incontestable advantage to noble metals such as silver or gold, in comparison with aluminum for instance, for which the intrinsic efficiency slightly exceeds 0.8 at $\lambda_0$ = 850 nm for both metals.

It is also interesting to consider molecules with a polarization parallel to the interfaces. The results are very different. Since the parallel electric-field component of the gap plasmon is much weaker than the perpendicular component, plasmonic modes are weakly excited for parallel polarizations, see Figs. 3(c-d), and quenching is now the dominant decay channel, even for tiny gap thicknesses. Likewise, since optical pumping of molecules placed in nanogaps is dominantly performed via the excitation of gap plasmons, optical pumps are much less efficient for molecules oriented parallel to the interfaces than for those oriented vertically.

All in all, one should not be afraid of tiny gaps to implement emitting optical devices, and nanogap antennas may really offer new opportunities not met with dielectric cavities. There is still a long way to go before optimizing the performance of plasmonic antennas, starting from designing nanogap geometries to match the impedance of slow gap plasmons with free-space photons to make the extrinsic efficiency $\eta_e$ approaching 100%. In this connection, $\eta_i$ appears as an upper bound for the antenna efficiency, which is ultimately limited by quenching and thanks to its great simplicity, Eq. (2) may provide inspiration to wisely select quantum emitters, dielectric insulators and plasmonic materials [17] in a specific wavelength range.

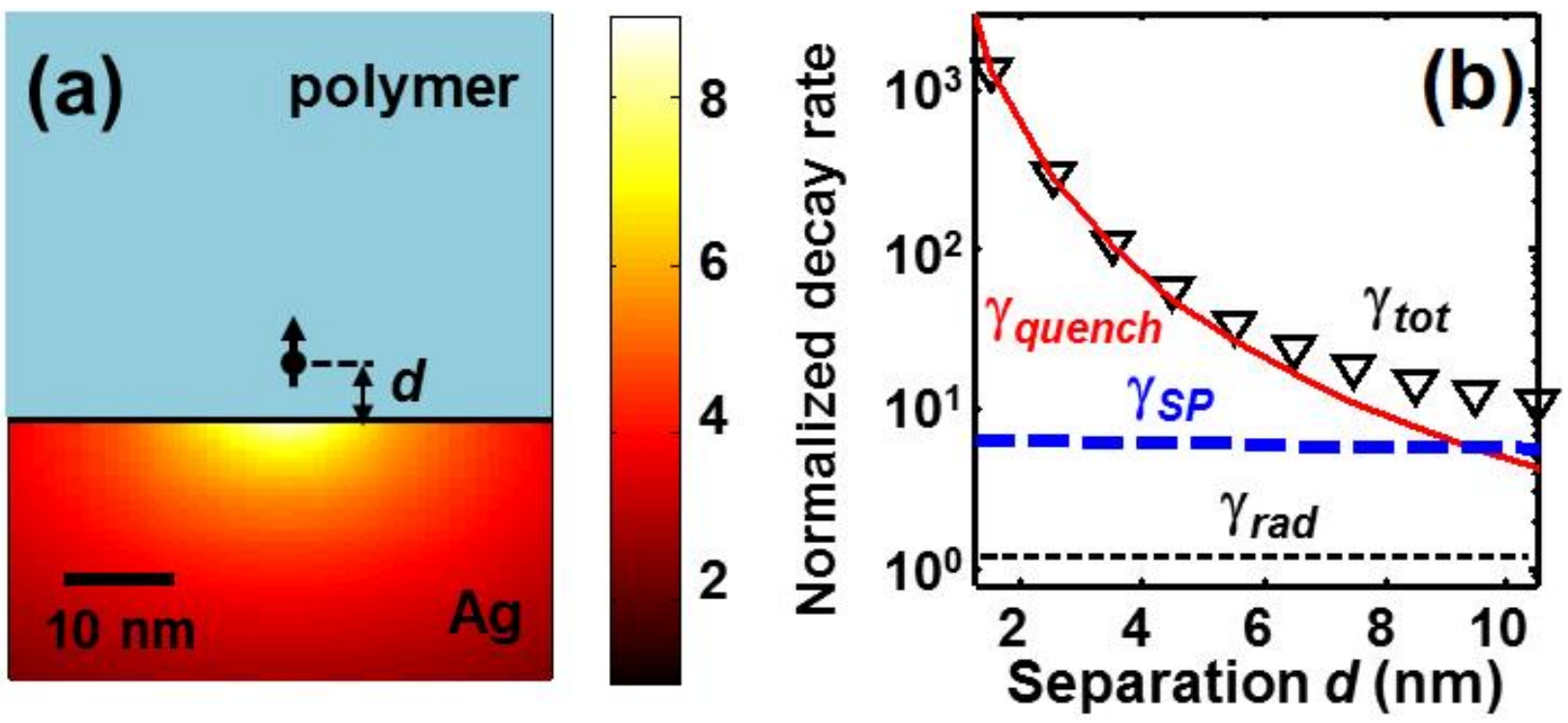

**Figure 1**. **Radiation of a vertical electric dipole above a Ag/polymer interface**. (**a**) $\log_{10}(|E|^2)$ of the total electric field excited by the dipole (black arrow) for a dipole-metal separation distance $d$ = 4 nm. (**b**) Calculated decay rates into SPPs ($\gamma_{SP}$, blue), free space photons ($\gamma_{rad}$, black), and quenching ($\gamma_{quench}$, red). The total decay rates $\gamma_{tot}$, is shown with black triangles. All decay rates are normalized by the decay rate in a vacuum. The calculations are performed for an emission wavelength $\lambda_0$ = 650 nm. The refractive index of polymer is $n$ = 1.4 and the silver permittivity is $\varepsilon_{Ag} = -17 + 1.15i$ .

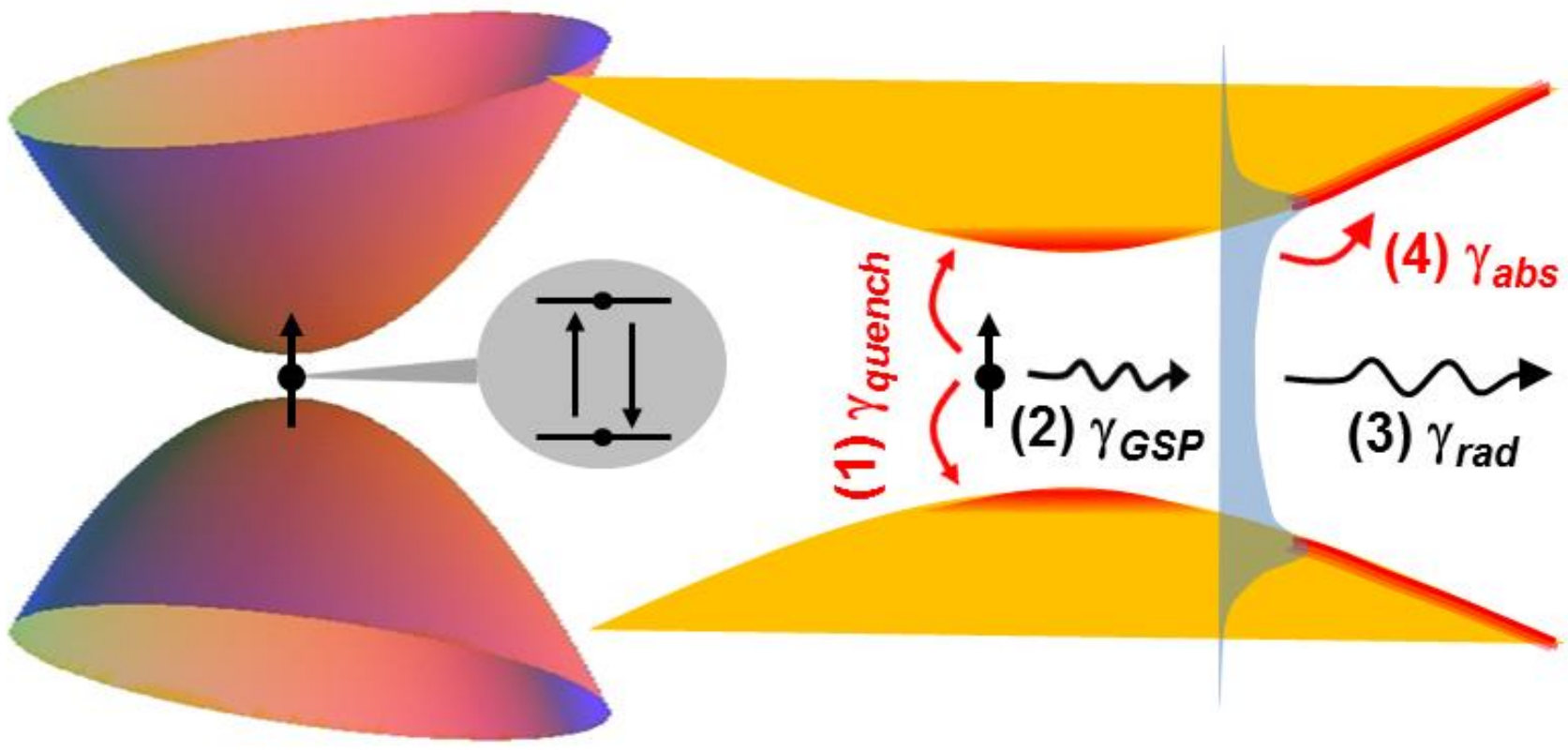

**Figure 2. Schematics for understanding modified emission in tiny gaps.** (1) Near-field non-radiative decay (quenching) at rate $\gamma_{quench}$ of the emitter into the metal; (2) Excitation of gap plasmons at rate $\gamma_{GSP}$; (3) Conversion of the excited plasmons into free space photons at rate $\gamma_{rad}$; (4) Plasmon decay into metal at rate $\gamma_{abs}$. The quantum emitter is assumed to have an 100% internal quantum efficiency.

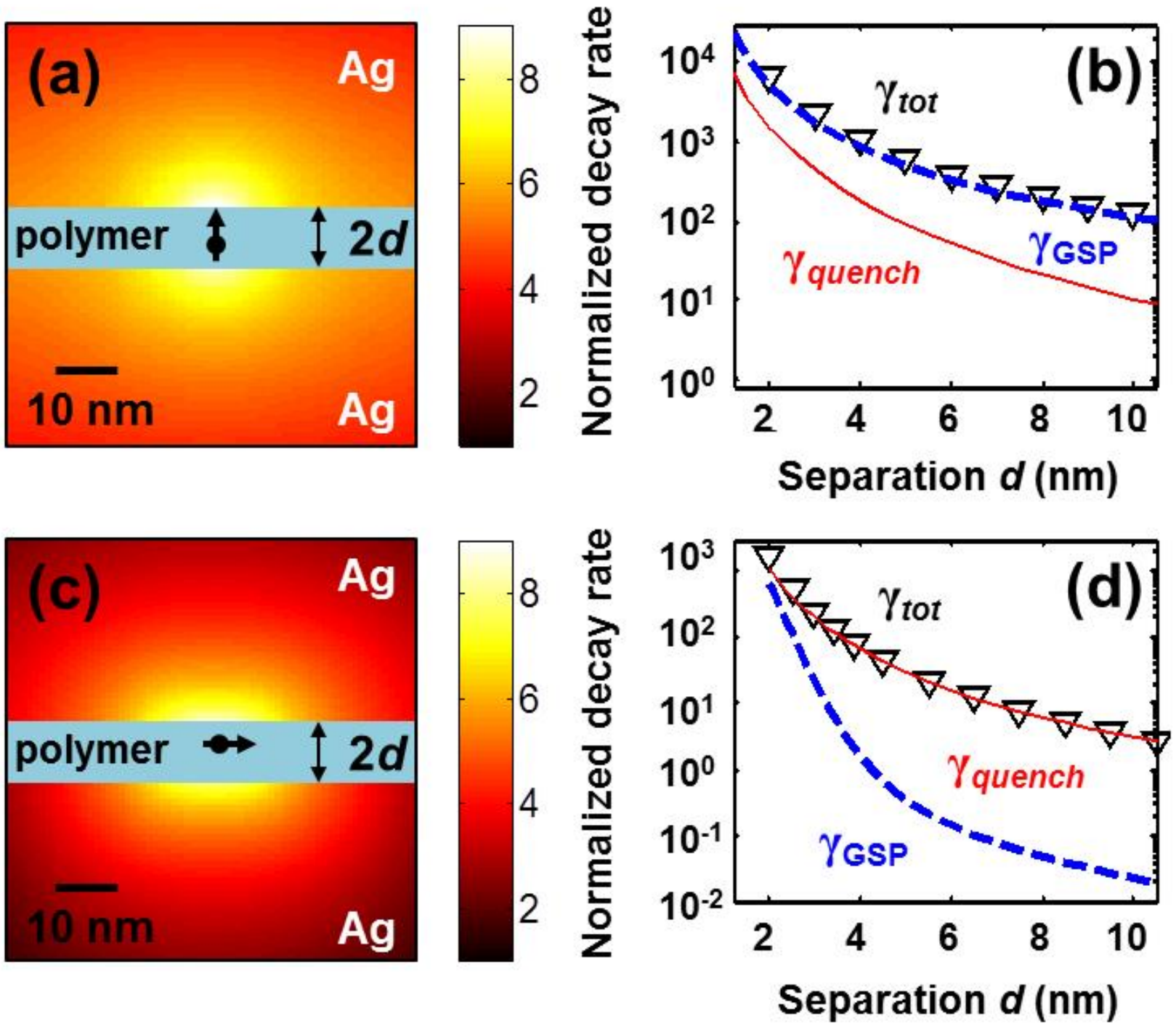

**Figure 3. Radiation of an electric dipole placed inside an Ag/polymer/Ag nanogap.** (**a**) and (**b**) vertical electric dipole placed at the center of the nanogap; (**c**) and (**d**) horizontal electric dipole that is placed 1 nm above the nanogap center (horizontal electric dipoles placed in the gap center do not couple to the gap plasmon mode). (**a**) and (**c**) $\log_{10}(|E|^2)$ of the total electric field radiated by the dipole (black arrow) for a gap thickness 2$d$ = 8 nm. (**b**) and (**d**) same legend as in Fig. 1b, except for the additional dashed-blue curves that represent the decay rates into gap plasmons.